\pdfoutput=1
\documentclass[aps,reprint,10pt,twocolumn,superscriptaddress]{revtex4-1}
\usepackage{amsmath, amssymb, amsthm, mathtools, bbm, physics}

\usepackage[utf8]{inputenc}
\usepackage{graphicx}
\usepackage{framed}
\usepackage{hyperref}
\usepackage{booktabs}
\usepackage{soul}

\usepackage{subcaption}
\usepackage{ragged2e} % for the \justifying macro
\DeclareCaptionJustification{justified}{\justifying}

\usepackage{microtype}
\microtypecontext{spacing=nonfrench}
\microtypesetup{
protrusion={true,nocompatibility},
activate={true,nocompatibility},
tracking=true,
kerning=true,
spacing={true}
}
  
\usepackage{tikz} % TikZ
\usetikzlibrary{calc, arrows, patterns, shapes, decorations, arrows.meta, fit, positioning}
\tikzset{
    auto,node distance =1 cm and 1 cm,semithick,
    var/.style = {minimum width = 0.5cm},
    intvar/.style = {circle, draw, minimum width = 0.5cm, double},
    latent/.style = {minimum width = 0.5cm},
    point/.style = {circle, draw, inner sep=0.06cm, fill, node contents={}},
    triangle/.style = {regular polygon, regular polygon sides=3, draw, inner sep=0.06cm, fill, node contents={}},
    bidir/.style={Latex-Latex,dashed},
    dir/.style={-Latex},
    el/.style = {inner sep=2pt, align=left, sloped}
}
\tikzstyle{vertex}=[circle, fill=black!10, draw=black]
\tikzstyle{edge}=[thick]
\tikzstyle{clique}=[line width=4, draw=black!70]

\graphicspath{{imgs/}}

\newcommand{\Do}[1]{\mathrm{do}(#1)}

\newcommand{\eye}{\mathbbm{1}}

\newcommand{\Cuc}{\mathcal{C}_\mathrm{UC}}
\definecolor{forestgreen}{rgb}{0.13, 0.55, 0.13}

\begin{document}
\title{A minimal example of quantum non-classicality without freedom of choice}

\author{Pedro Lauand}
\affiliation{Instituto de Física Gleb Wataghin, Universidade Estadual de Campinas, 13083-859, Campinas, Brazil}
\author{Davide Poderini}
\affiliation{International Institute of Physics, Federal University of Rio Grande do Norte, 59078-970, Natal, Brazil}
\author{Rafael Rabelo}\textit{}
\affiliation{Instituto de Física Gleb Wataghin, Universidade Estadual de Campinas, 13083-859, Campinas, Brazil}
\author{Rafael Chaves}
\affiliation{International Institute of Physics, Federal University of Rio Grande do Norte, 59078-970, Natal, Brazil}
\affiliation{School of Science and Technology, Federal University of Rio Grande do Norte, Natal, Brazil}

\begin{abstract}
Bell's theorem is often considered the most stringent notion of non-classicality. The generalization of Bell's theorem to causal networks offers interesting new perspectives on the phenomenon of quantum non-classicality and prompts us with a fundamental inquiry: What is the simplest scenario leading to the incompatibility between quantum correlations and the classical theory of causality? Here we demonstrate that quantum non-classicality is possible in an entanglement swapping network consisting of only three dichotomic variables, without the need for the locality assumption or external measurement choices. We also show that interventions, a central tool in the field of causal inference, significantly improve the noise robustness of this new kind of non-classical behavior, making it feasible for experimental tests with current technology.
\end{abstract}

\maketitle
\emph{Introduction.---} Bell’s theorem \cite{bell1964einstein} stands as a cornerstone in the foundations of quantum theory, revealing that certain correlations arising from entangled quantum systems cannot be explained by any classical causal model satisfying locality and realism. This insight led to the development of the device-independent paradigm \cite{brunner2014bell} and inspired generalizations to causal networks, uncovering new and stronger forms of non-classicality \cite{tavakoli2022bell}.

From the perspective of causality theory \cite{pearl2009causality}, Bell’s theorem can be framed as a causal compatibility problem \cite{wood2015lesson,chaves2015unifying,wolfe2019inflation}: determining whether a given classical causal structure can explain the observed data. Network generalizations introduce two key new ingredients: independent sources that mediate correlations between network nodes \cite{branciard2010characterizing,branciard2012bilocal}, and time-like causal influences, where earlier outcomes can affect later measurements \cite{chaves2018quantum,brask2017bell}. While each has independently led to novel insights, such as the necessity of complex numbers in quantum theory \cite{renou2021quantum}, refined notions of multipartite non-locality \cite{coiteux2021no,suprano2022experimental,pozas2022full}, self-testing of physical theories \cite{weilenmann2020self} and new non-classicality witnesses based on causal bounds  \cite{ringbauer2016experimental,gachechiladze2020quantifying,agresti2022experimental,poderini2024observational}, the combination of these new ingredients remains largely unexplored.

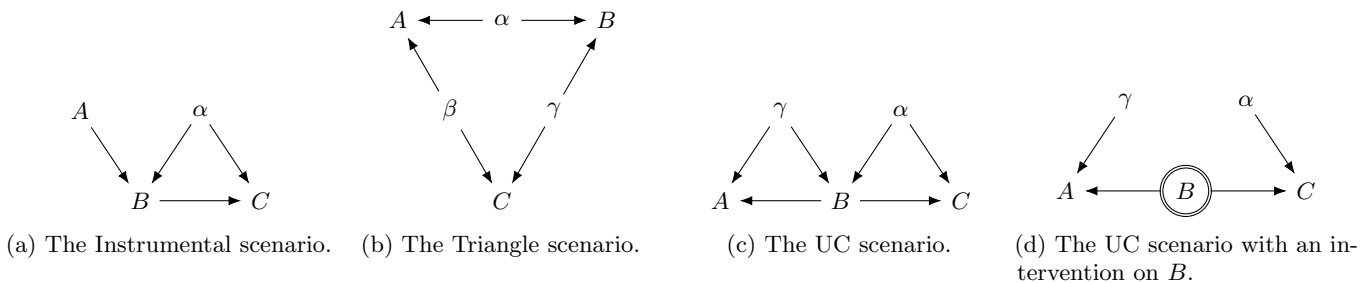
\begin{figure*}
\begin{subfigure}[t]{.24\textwidth}
\begin{tikzpicture}
    \node[var] (b) at (1.6,0) {$C$};
    \node[var] (a) at (0,0) {$B$};
    \node[var] (x) at (-.8,1.2) {$A$};
    \node[latent] (l) at (.8,1.2) {$\alpha$};
    % Directed edges
    \path[dir] (x) edge (a) (a) edge (b); 
    \path[dir] (l) edge (a) (l) edge (b);
\end{tikzpicture}
\caption{The Instrumental scenario.}
\label{fig:instrumental_dag}
\end{subfigure}
\begin{subfigure}[t]{.24\textwidth}
    \begin{tikzpicture}

        \foreach [count=\k] \l/\n/\a in {1/A/\gamma, 2/C/\beta, 3/B/\alpha} {
            \draw (\k*360/3 - 30: .8cm) node[latent] (l\k) {$\a$};
            \draw (\k*360/3 + 30: 1.6cm) node[var] (\n) {$\n$};
			\path[dir] (l\k) edge (\n);
        }
		\foreach \k/\l in {1/B, 2/A, 3/C}
			\path[dir] (l\k) edge (\l);

    \end{tikzpicture}
    \caption{The Triangle scenario.}
    \label{fig:triangle_dag}
    \end{subfigure}
\begin{subfigure}[t]{.25\textwidth}
\begin{tikzpicture}
    \node[var] (a) at (-1.6,0) {$A$};
    \node[var] (c) at (1.6,0) {$C$};
    \node[var] (b) at (0,0) {$B$};
    \node[latent] (l) at (-.8,1.2) {$\gamma$};
    \node[latent] (g) at (.8,1.2) {$\alpha$};
    % Directed edges
    \path[dir] (l) edge (a) (l) edge (b); 
    \path[dir] (g) edge (b) (g) edge (c); 
    \path[dir] (b) edge (a) (b) edge (c);
\end{tikzpicture}
    \centering        
    \caption{The UC scenario.}
    \label{fig:evans_dag}
\end{subfigure}
\begin{subfigure}[t]{.25\textwidth}
\begin{tikzpicture}
    \node[var] (a) at (-1.6,0) {$A$};
    \node[var] (c) at (1.6,0) {$C$};
    \node[intvar] (b) at (0,0) {$B$};
    \node[latent] (l) at (-.8,1.2) {$\gamma$};
    \node[latent] (g) at (.8,1.2) {$\alpha$};
    % Directed edges
    \path[dir] (l) edge (a); 
    \path[dir] (g) edge (c); 
    \path[dir] (b) edge (a) (b) edge (c);
\end{tikzpicture}
    \centering        
    \caption{The UC scenario with an intervention on $B$.}
    \label{fig:evansint_dag}
\end{subfigure}
\caption{\textbf{All non-trivial causal structures with three-observable nodes}. a) The instrumental scenario features the relaxation of locality. b) The triangle scenario includes independent sources of correlations and does not require external inputs. c) The UC causal structure merges the core features of the instrumental and triangle DAGs: relaxation of locality, source independence, and the absence of external inputs. d) UC scenario with interventions made on the central node of the causal network. Non-classicality here means the possibility of a mismatch between classical and quantum predictions, which, as proved here, indeed happens for all of them.
}
\label{fig:all_dags}
\end{figure*}

One prominent question arising in this broader context is that of minimal quantum non-classicality, where several notions of minimality can be identified. First, in terms of observable variables: while the standard Bell scenario involves four, some causal structures, like the triangle \cite{fritz2012beyond,renou2019genuine,vsupic2020quantum,chaves2021causal,polino2023experimental}, instrumental \cite{pearl1995testability,chaves2018quantum,VanHimbeeck2019quantumviolationsin}, and Unrelated Confounders (UC) scenarios \cite{lauand2023witnessing,camillo2024estimating}, require only three observable variables \cite{evans2016graphs} (see Fig. \ref{fig:all_dags}). Second, minimality can refer to the cardinality of variables: as we show here, the UC scenario displays quantum non-classicality with binary variables, whereas the triangle~\cite{fraser2018triangle,boreiri2023towards,pozas2023post} and instrumental scenarios \cite{henson2014theory} do not. Third, minimality may refer to the dimensionality of the observed probability space, and once again, the UC case leads to the minimum number of parameters. Finally, minimality may concern causal assumptions: whereas Bell’s theorem requires both locality and freedom of choice, the instrumental and triangle scenarios each relax one of them; the UC scenario relaxes both simultaneously.

Despite the foundational significance of the UC scenario, which furthermore underlies entanglement swapping experiments \cite{zukowski1993event}, its quantum non-classicality has remained elusive. Previous results showed that large classes of quantum strategies, even involving maximally entangled states, yield correlations that admit classical explanations \cite{lauand2023witnessing}. Post-quantum violations were found \cite{camillo2024estimating}, but no such violations had been shown for quantum theory.

In this work, we develop a data-driven framework based on quadratic programming to derive polynomial Bell inequalities and employ it to demonstrate quantum correlations that violate the classical constraints of the UC scenario using only three dichotomic variables. This result has three main implications. First, it proves that non-classical quantum correlations can arise in the UC scenario, even when known families of entangled states fail to do so. Second, it shows that all causal structures leading to Bell-like inequalities with three observable variables admit quantum violations. And finally, it establishes the UC scenario as a minimal example of quantum non-classicality according to all notions of minimality discussed above: number of observable nodes, variable cardinality, dimensionality of the probability space, and causal assumptions. Furthermore, we show that the use of interventional data \cite{gachechiladze2020quantifying} significantly enhances the noise robustness of quantum violations, making the UC scenario arguably the most experimentally feasible network for observing quantum non-classicality beyond Bell's theorem.

\emph{Unrelated Confounders Scenario---}The UC scenario \cite{evans2016graphs,lauand2023witnessing} (see Fig. \ref{fig:evans_dag}) is, notably, the simplest network combining two fundamental ingredients -- source independence and communication -- which does not require external inputs. However, it remains an open question whether quantum non-classicality holds for the UC case \cite{lauand2023witnessing,camillo2024estimating}. This network features three observed variables $A$, $B$, and $C$, for the outcomes of Alice, Bob, and Charlie, respectively; and independent sources of correlation, $\gamma$, between $AB$, and $\alpha$, between $BC$. Classically, observations $P(a,b,c)$ compatible with such causal structure factorize as 
\begin{equation}
\label{eq:clas}
P(a,b,c)=\sum_{\alpha,\gamma}P(\alpha)P(\gamma)P(a|b,\gamma)P(b|\alpha,\gamma)P(c|b,\alpha)
\end{equation}
such that $P(\alpha,\gamma)=P(\alpha)P(\gamma)$. For quantum theory, we have the sources distributing quantum information; the global state is $\psi_{ABC}=\psi_\gamma \otimes \psi_{\alpha}$ and the POVMs describing the measurements on it are given by $E_{a|b}$, $E_{b}$ and $E_{c|b}$. Born's rule implies, then, that the observed probabilities obtained when we impose the UC causal structure on a given experiment are given by
\begin{equation}
\label{eq:quan}
    P(a,b,c)=\Tr \left(\psi_\gamma \otimes \psi_{\alpha}\left( E_{a|b}\otimes E_b \otimes E_{c|b}\right)\right).
\end{equation}

As will be shown in the following, in order to reveal the non-classicality in the UC scenario, it is enough to consider $a,b,c \in \{-1,1\}$. Since $B$ has access to all the information from the sources $\gamma$ and $\alpha$, it can send information from $\alpha$ to $A$ and, similarly, from $\gamma$ to $C$. However, this forwarding of information is limited by the cardinality of $B$. This suggests to look at the correlation between $A$ and $C$ for each value of $b$ and, therefore, we consider \emph{correlators}, given by
\begin{equation}
  \langle A^{i}C^{j}\rangle_b:= \sum_{a,c}a^i c^ j P(a,b,c),
\end{equation}
for $i,j\in \{0,1\}$ and also including Bob's correlator $\langle B \rangle  = \sum_b  bP(b)$. Notably, the correlators have a one-to-one correspondence with $P(a,b,c)$.

\emph{Quantum strategy ---} To illustrate our approach, we will consider a specific quantum implementation of the UC experiment. We let both sources emit pairs of maximally entangled qubits, $|\psi^{+}\rangle=\frac{1}{\sqrt{2}}(|01\rangle+|10\rangle)$, and let Bob perform a joint entangling projective measurement $\{|\psi_{\theta}\rangle \langle \psi_{\theta}|, \eye-|\psi_{\theta}\rangle \langle \psi_{\theta}| \}$, where Bob outputs $b=1$ if he can successfully project his subsystems in the state $|\psi_{\theta}\rangle $, given by 
\begin{equation}
    |\psi_{\theta}\rangle =\sin(\theta)|01\rangle + \cos(\theta)|10\rangle,
    \label{eq:projB}
\end{equation}
and outputs $b=-1$ otherwise. If the projection is successful, Bob effectively prepares Alice’s and Charlie’s joint state in an entangled state, given by
\begin{equation}
    |\varphi_{\theta}\rangle =\cos(\theta)|01\rangle + \sin(\theta)|10\rangle,
\end{equation}
and communicates his outcome to Alice and Charlie. Upon receiving Bob's outcome, if the projection is successful, Alice and Charlie perform projective measurements on the $\sigma_x$ basis or, if $b=-1$, they measure their ``trash" state in the computational basis $\sigma_z$ and output $a,c\in\{-1,1\}$. Quantum theory predicts that the observed probability distribution is 
\begin{equation}
\label{eq:quantump}
\begin{aligned}
     &P_Q(a,b=1,c)=\frac{1}{16}\left(1+\sin(2\theta)ac\right),\\
     &P_Q(a,b=-1,c)=\frac{1}{16}\left(3+ ac+\cos(2\theta)\left(c-a\right)\right),
\end{aligned}
\end{equation}
with corresponding correlators given by
\begin{equation}
    \label{eq: quantum_corr}
\begin{aligned}
     &\langle AC \rangle_{1}= \frac{\sin(2\theta)}{4}\\
     &\langle A \rangle_1=\langle C \rangle_1=0\\
     &\langle C \rangle _{-1}=-\langle A \rangle_{-1}=\frac{\cos(2\theta)}{4},\\
     &\langle AC \rangle_{-1}=\frac{1}{4}\\
     &\langle B \rangle = -\frac{1}{2}.
\end{aligned}
\end{equation}

\emph{Data-driven inequalities---} 
Systematic methods for deriving non-linear inequality constraints in networks are well-documented \cite{chaves2016polynomial,tavakoli2022bell}. When dealing with causal networks involving multiple independent sources, the classical set of correlations is semi-algebraic -- meaning it can be described by a finite number of polynomial Bell-like inequalities -- and generally non-convex \cite{garcia2005algebraic}. Existing techniques for deriving such network Bell inequalities rely on variable elimination methods from algebraic geometry \cite{geiger1999quantifier}. However, given their NP-hard computational complexity \cite{geiger1999quantifier}, their application to network scenarios is intractable even for the simplest possible cases \cite{ver2011sequence}.  Other systematic approaches have also leveraged the Hopcroft-Karp algorithm \cite{Hopcroft_algorithm}, although only characterizing specific families of Bell-like inequalities for general networks \cite{Ming-Xing_PRL}. Notably, no systematic full characterization of a network with more than one source has ever been proposed. To overcome such difficulties, \emph{data-driven} approaches to derive Bell-like inequalities for specific observed correlations have been well-established in the literature \cite{wolfe2019inflation,wolfe2021quantum,pozas2022full,pozas2023post}. A prominent example is the inflation technique \cite{wolfe2019inflation}, which allows us to relax the original non-convex problem to a linear program and derive Bell-like inequalities tailored to a specific correlation from convex duality optimization theory. However, the inflation technique becomes computationally intractable very fast and has been shown to perform poorly on the UC case \cite{camillo2024estimating}. Other data-driven approaches, such as \cite{armin_EJM_bilocal}, employed concave expressions to derive Bell-like inequalities. However, the validity of the inequalities either relied on symmetry ansätze of the local model or conjectured local bounds.

Here, going beyond the linear programming and convex optimization paradigm, we turn to global quadratic programming \cite{nocedal2006quadratic} to employ nonlinear expressions and capture the non-convexity of the set of classical correlations. A quadratic program (QP) can be understood as a generalization of a linear programming (LP) optimization problem, where the objective function and constraints are allowed to be quadratic. Notably, QP's are, in general, non-convex and a common approach for \emph{global} optimization of such problems is the \emph{branch and bound} method. This method works as follows: the non-convex QP is converted into a convex relaxation (e.g., an LP) to get a lower bound on the optimal solution and, if the solution of this relaxation does not satisfy the original non-convex constraints, the feasible region is split into smaller subproblems (a.k.a. branches). Then, it computes upper bounds by evaluating feasible solutions and, if a branch’s lower bound exceeds the best known upper bound, it is discarded. This procedure is repeated systematically until the difference between the lower bounds and upper bounds is very small, and this difference is usually taken to be the computational precision of the solver. This ensures a global solution of the original non-convex problem (up to computational precision). It is also possible to model general polynomial optimization with a QP using extra auxiliary variables and, in principle, adapt it as a data-driven approach for any network \cite{lauand2023witnessing}. To deal with such QP problems, we use the Gurobi optimizer \cite{gurobi}. 

To showcase the power of our approach, we use quadratic programming to obtain a description of the subspace of probability distributions inspired by our quantum protocol. That is, we consider the two-dimensional subspace $(E_{+},E_{-})$ of probability distributions given by:
\begin{equation}
\begin{aligned}
     &\langle AC \rangle_{1}= E_{+},\\
     &\langle A \rangle_1=\langle C \rangle_1=0,\\
     &\langle C \rangle _{-1}=-\langle A \rangle_{-1}=E_{-},\\
     &\langle AC \rangle_{-1}=\frac{1}{4},\\
     &\langle B \rangle = -\frac{1}{2}.
\end{aligned}
\end{equation}
where $|E_{i}|\leq \frac{1}{4}$ for $i=\pm$. Defining $I_{+} = 1+4E_{+}$ and $I_{-}=\frac{1+4E_{-}}{8}$ we obtain the network inequality
\begin{equation}
\label{eq: inequality_I}  \mathcal{I}=\sqrt{I_{+}}+3\sqrt{I_{-}}\leq C_{\mathcal{I}},
\end{equation}
with the bound $C_{\mathcal{I}}= \frac{3}{\sqrt2} + \frac{\sqrt{7}}{6}$ for classical models. We heuristically chose the coefficients in the square root terms, aiming to maximize the gap between the numerical value of our quantum strategy and the classical bound using QP. Then, solving these problems gives us optimal classical strategies, which are solutions to the QP, and then we approximate from above to finally obtain the upper bound presented. We find this inequality describes (up to relabeling symmetries) an accurate outer approximation of the set of classical correlations within this subspace $E_{\pm}$  (see Supplemental Material (SM) \cite{supp}). For our quantum correlation of Eq. \eqref{eq: quantum_corr}, we straightforwardly obtain $(I_{+},I_{-})=(1+\sin(2\theta), \frac{1+\cos(2\theta)}{8})$, and $\mathcal{I}(\theta)=\sqrt{1+\sin(2\theta)} + \frac{3}{2\sqrt{2}}\sqrt{1+\cos(2\theta)}$. As expected from previous results, the choice of $\theta=\frac{\pi}{4}$, i.e., when Bob's measurement project onto the maximally entangled state, we have $\mathcal{I}(\frac{\pi}{4})=\sqrt{2}+\frac{3}{2\sqrt{2}}\approx 2.47487 %2.474874 
< \frac{3}{\sqrt2} + \frac{\sqrt{7}}{6} \approx 2.56228 %2.562278895
$. However, by choosing $\theta=\pi/8$, i.e., by having $B$ measure in a partially entangled Bell basis, we find $\mathcal{I}(\frac{\pi}{8}) = \sqrt{2+\sqrt{2}}\left(\frac{3\sqrt{2}+4}{4\sqrt{2}}\right)\approx 2.69238 > 2.56228$.

\begin{center}
    \begin{figure}
    \centering
    \includegraphics[width=\columnwidth]{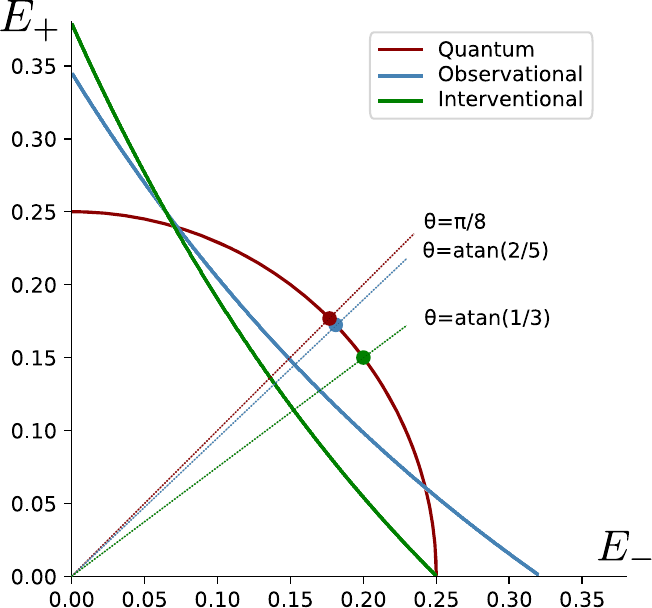}
    \caption{\textbf{Maximum violations in the $E_{\pm}$ subspace.} Allowed regions of the subspace parametrized by $E_+$ and $E_-$. 
    The blue and green curves represent, respectively, the classical boundaries described by the observational inequality~\eqref{eq: inequality_I} and the interventional inequality~\eqref{eq: inequality_F}, while the red curve represents the values given by the quantum distribution~\eqref{eq:quantump}. 
    The points show the maximum violations attainable in each case (see SM \cite{supp}), along with the point $\theta=\pi/8$ corresponding to the strategy presented here.}
    \label{fig:subspace}
\end{figure}
\end{center}

\emph{Interventions---}Importantly, the presence of communication of the output of $B$ to $A$ and $C$ allows us to go beyond passive observations of the experiment and investigate the role of interventions \cite{balke1997bounds,pearl2009causality}. In contrast to observations, interventions allow one to locally change the causal structure, eliminating all external influences of a given variable $X$, forcing it to a particular value $x$, denoted by $do(X=x)$, while keeping the rest of the model unchanged.  Ultimately, this consists of collecting additional statistics under different experimental conditions. Classically, an intervention in node $B$ of the UC scenario (see Fig. \ref{fig:evansint_dag}) breaks the correlations between nodes $A$ and $C$, and its statistics define a do-conditional given by
\begin{equation}
\label{eq:dob}
    P(a,c|\Do{b})=P(a|\Do{b})P(c|\Do{b}),
\end{equation}
where $P(a|\Do{b})=\sum_{\gamma}P(\gamma)P(a|b,\gamma)$ and similarly for $P(c|\Do{b})$. In turn, if the sources distribute quantum states, the do-conditional still respects \eqref{eq:dob}, with
\begin{equation}
P(a|\Do{b})=\Tr\left(\psi_{\gamma} \; E_{a|b}\otimes\eye \right),
\end{equation}
and similarly for $P(c|\Do{b})$.
Following the same reasoning as in the last sections, we can add interventional data to our analysis. Defining the correlators $\langle A \rangle_{\Do{b}}=\sum_a aP(a|\Do{b})$ and $\langle C \rangle_{\Do{b}}=\sum_c cP(c|\Do{b})$, by direct computation our protocol yields
\begin{equation}
    \langle A \rangle_{\Do{b}}=\langle C \rangle_{\Do{b}}=0,
\end{equation}
and we can directly incorporate these extra constraints into the $(E_+,E_-)$ subspace. In this new subspace, we propose the Bell expression
\begin{equation}
\label{eq: inequality_F}
    \mathcal{F}=\sqrt{I_{+}}+4\sqrt{I_{-}} \leq C_{\mathcal{F}},
\end{equation}
with $C_{\mathcal{F}}=\frac{9}{7} + \frac{7}{10}\sqrt{6}$. The coefficients inside the square root are chosen to maximize the gap between the classical bound $C_{\mathcal{F}}$ and its maximum quantum violation (see details in SM \cite{supp}). For our quantum strategy, we obtain $\mathcal{F}(\theta)=\sqrt{1+\sin(2\theta)}+\sqrt{2}\sqrt{1+\cos(2\theta)}$ and, again, the choice $\theta=\frac{\pi}{8}$ yields $\mathcal{F}(\frac{\pi}{8})=\frac{1}{2} (2 + \sqrt{2})^{\frac{3}{2}} \approx 3.15432 > \frac{9}{7} + \frac{7}{10}\sqrt{6} \approx 3.00036$.

\emph{Beyond the $E_{\pm}$-subspace---} We have shown how one can find quantum non-classical correlations restricted to the $E_{\pm}$-subspace. However, in practice, the sources and measurement apparatuses are always subjected to imperfections, which would yield observations that slightly deviate from the subspace we are studying. To address this, we generalize the inequalities $\mathcal{I}$ and $\mathcal{F}$ beyond the $E_{\pm}$-subspace and derive robust witnesses for non-classicality. We give a brief description here of these generalized inequalities and defer the details to the SM \cite{supp}. 

For the observational case, this is done by incorporating penalty functions that attempt to guarantee all the constraints from the $E_{\pm}$-subspace. Specifically, we consider the probability distribution $P(a,b,c)$ and the penalty function 
\begin{align*}
    s(P)=&18\left|P(b=1)-\frac{1}{4}\right|+18\left|\sum_{a\neq c}P(a,-1,c)-\frac{1}{4}\right|\\
    &+4\sum_{a=c}\left|P(a,-1,c)-\frac{1}{4}\right| +4\left|\langle AC\rangle_{-1}-\frac{1}{4}\right| \\
    &+|\langle A\rangle_{-1}+\langle C\rangle_{-1}|+|\langle A\rangle_{1}|+|\langle C\rangle_1|.
\end{align*}
Using this penalty function and QP, one can prove that the inequality
\begin{equation}
\label{eq:ineqI}
    \mathcal{I'}= 2\sum_{a= c}\sqrt{P(a,1,c)}+3\sqrt{P(-1,-1,1)} -s(P)\leq C_{\mathcal{I'}},
\end{equation}
with $C_{\mathcal{I'}}=\frac{11}{6}+\frac{4\sqrt{3}}{9}\approx 2.60313$, is valid for any probability distribution compatible with the classical UC causal network, with $\mathcal{I'}$ reducing to $\mathcal{I}$ into the subspace of $E_+$ and $E_-$ (see SM \cite{supp}).

To probe this non-classicality, we consider a noise model that introduces isotropic noise with a visibility $v$ to the quantum sources. For an imperfect Bell-state measurement, we introduce an efficiency parameter $\eta$ such that the success probability of the measurement is $\frac{1+\eta}{2}$. For $v=1$, we can find violation of $\mathcal{I'}$ for any $\eta \geq 0.69$. Taking $v = 0.99$, we can find violation of $\mathcal{I'}$ for $\eta \geq 0.97$. The best critical visibility we could find is for $v_{\text{crit}} \approx 0.98864$, below which there is no violation of the inequality. (See details in the SM \cite{supp}).

For the interventional case, we generalize the inequality $\mathcal{F}$ to $\mathcal{F'}$, which, as we will see, allows for a much more experimentally feasible test of non-classicality. For that, we must incorporate \emph{do-conditional} probabilities into the penalty function of $\mathcal{F'}$. Here, we propose
\begin{align*}
    g(P)=&\sum_{a=c}\left|P(a,-1,c)-\frac{1}{4}\right|
        +\left|\langle AC\rangle_{-1}-\frac{1}{4}\right|\\
       & + \left|\sum_{a\neq c}P(a,-1,c)-\frac{1}{4}\right|
        +|\langle A\rangle_{-1}+\langle C\rangle_{-1}|\\
   & +|\langle A\rangle_1|+|\langle C\rangle_1|
        +18\left|P(b=1)-\frac{1}{4}\right|\\
   & +\sum_b\left(|\langle A\rangle_{\Do{b}}|+|\langle C\rangle_{\Do{b}}|\right).
\end{align*}
which leads to hybrid observational-interventional non-linear inequality bounding any classical correlations in the UC scenario, and given by 
\begin{equation}
    \mathcal{F'}=2\sum_{a= c}\sqrt{P(a,1,c)}+4\sqrt{P(-1,-1,1)} -g(P) \leq C_{\mathcal{F}}.
\end{equation}
Introducing the same noise models as before, with a visibility $v$ for the sources and efficiency parameter $\eta$ for the entangled measurement, we find that for $v = 1 $ the bound is violated for any $\eta \geq 0.5$; and for visibilities as low as $v = 0.9$ we still find violations for $\eta \geq 0.85$. The critical visibility we could find was $v_{\mathrm{crit}} \approx 0.8774$, bringing the test of this new form of non-classicality within the reach of current technology (for a more in-depth analysis of the robustness to noise, see SM \cite{supp}).

\emph{Discussion.---} The simplest causal networks that can lead to quantum correlations without a classical explanation require at least three observable variables (notice that a standard bipartite Bell scenario has four observable variables: the choices of measurements and the corresponding outcomes). Within this context, it is known that there are only three different classes of DAGs \cite{evans2016graphs}: the instrumental \cite{pearl1995testability,chaves2018quantum}, the triangle \cite{fritz2012beyond,renou2019genuine}, and the unrelated confounders \cite{lauand2023witnessing} scenarios, shown in Fig.~\ref{fig:all_dags}. It was known that the first two can lead to quantum non-classicality \cite{chaves2018quantum,fritz2012beyond}, and here we prove that the same holds true for UC. Since all three observable node networks can lead to quantum violations of causal constraints, a natural question is to understand how generic quantum non-classicality is. Does it happen in any non-trivial causal network? If not, what are the key ingredients for the emergence of non-classicality? As previous results have shown and our present findings reinforce, locality and freedom of choice, unavoidable requirements in a standard Bell test, can be relaxed.

The instrumental case partially relaxes the locality assumption in Bell's theorem, since the measurement outcome of one observer can define the measurement basis for the other. However, one still has to employ an external random variable as an input and invoke the instrumentality assumption that the measurement choice of one observer has no direct causal influence over the other. In turn, the triangle scenario allows for non-classicality without the need for external inputs. However, such non-classicality requires a space-like separation between all three parties in the network and three independent entanglement sources as well as three entangled measurements to be tested experimentally \cite{wang2024experimental}.

The UC scenario we analyze here allows for much more experimentally feasible tests of non-classicality in quantum networks, requiring only two sources of entanglement and one entangled measurement. Specifically, the quantum strategy we propose is based on a two-outcome incomplete Bell state measurement (achievable
with linear optics), in stark contrast with known proposals for the triangle network~\cite{renou2019genuine,boreiri2023towards} and the bilocal network~\cite{bej2022creating}. The scenario does not feature any external inputs and at the same time provides a strong relaxation of the locality assumption, since one party explicitly communicates its measurement outcomes. Furthermore, such quantum violations of classical constraints in this causal structure are possible in the minimal case where all observable nodes are dichotomic, something that is not possible in the instrumental~\cite{henson2014theory} case and is still an open problem for the triangle network~\cite{boreiri2023towards,pozas2023post}. Finally, even though in the purely observational case the required visibilities for the two sources of entangled states are very high, the use of \emph{interventions} can bring them down to values that are certainly amenable to current technologies, also showing the power of interventional data as witnesses of non-classicality. 

Moving beyond the foundational aspects, a practical and relevant direction is to investigate the quantum information processing capabilities of the UC network.
In particular, the absence of external inputs makes it extremely interesting for randomness certification~\cite{sekatski2023partial, minati2025experimental} and other cryptographic and communication protocols~\cite{lee2018towards}.

\begin{acknowledgments}
We thank Robert Spekkens for helpful discussions in the early steps of this project. This work was supported by the Serrapilheira Institute (Grant No. Serra-1708-15763), the Simons Foundation (Grant Number 1023171, RC), the Financiadora de Estudos e Projetos (grant 1699/24 IIF-FINEP),
São Paulo Research Foundation FAPESP (Grant No. 2022/03792-4), the Brazilian National Council for Scientific and Technological Development (CNPq) (INCT-IQ and Grant Nos. 307295/2020-6, 316657/2023-9 and No.403181/2024-0).
\end{acknowledgments}

\bibliography{biblio}

\clearpage
\onecolumngrid
\begin{center}
{\large\bfseries Supplemental Material for\\[0.3em] ``A minimal example of quantum non-classicality without freedom of choice''}\\[1em]
Pedro Lauand,$^{1}$ Davide Poderini,$^{2}$ Rafael Rabelo,$^{1}$ and Rafael Chaves$^{2,3}$\\[0.5em]
{\small\itshape
$^{1}$Instituto de F\'isica Gleb Wataghin, Universidade Estadual de Campinas, 13083-859, Campinas, Brazil\\
$^{2}$International Institute of Physics, Federal University of Rio Grande do Norte, 59078-970, Natal, Brazil\\
$^{3}$School of Science and Technology, Federal University of Rio Grande do Norte, Natal, Brazil}
\end{center}
\twocolumngrid

\setcounter{equation}{0}
\renewcommand{\theequation}{S\arabic{equation}}
\setcounter{section}{0}
\renewcommand{\thesection}{S\Roman{section}}

\section{Inequalities beyond the $E_{\pm}$ subspace} 

Causal networks with independent sources, as is the case in the UC scenario, have a non-convex set of correlations. In the main text, we described how we can obtain a classical bound for inequalities $\mathcal{I}$ and $\mathcal{F}$ by describing the problem as an instance of Quadratic Programming (QP) optimization and carefully analyzing its solution and obtaining suitable upper bounds for the inequalities. However, our inequalities are only valid for the $E_{\pm}$ subspace and are not suited to analyze noise models. Here, we show how one can leverage quadratic programming (QP) to circumvent these issues and derive useful data-driven and robust non-linear witness for non-classicality. For completeness, we will show the explicit formulation of such QP problems.

Trying to capture the independence of the two sources represented by the quadratic constraint $P(\alpha,\gamma)=P(\alpha)P(\gamma)$, inspired by previous proposals \cite{branciard2010characterizing,tavakoli2014nonlocal,chaves2016polynomial} we considered Bell inequalities which may be expressed by a functional depending on the square root of the probability $P(a,b,c)$ of the experiment, i.e. $f(\sqrt{P(a,b,c)})$. Furthermore, to wave the $E_{\pm}$ subspace constraints and obtain inequalities valid for all the set of classical correlations in the UC network, we can also add penalty functions depending on the absolute value of some function of $P(a,b,c)$. For instance, suppose that in a given experiment, we observe that Alice has random outcomes, i.e., $P(a)=\frac{1}{2}$. Adding a term like $\vert P(a) -\frac{1}{2} \vert$ to the inequality, we construct a witness tailored to that specific observed correlation. Thus, in the following, we will focus on inequalities with the following functional form
\begin{equation}
\label{eq:ineqgen}
   \min_{x,y} \sum_i |x_i|+\sum_i b_i\sqrt{y_i},
\end{equation}
where $x_i$ and $y_i$ are linear functions of $P(a,b,c)$.
Given a certain expression of the form \eqref{eq:ineqgen}, our aim is to find its maximum value in a classical description of causality and, even though we are dealing with a nonconvex optimization problem, the general framework of linear programming (LP) \cite{boyd2004convex} can still be quite versatile and useful for our purposes. For instance, we can optimize expressions like $|x|$ by defining a new variable $x'$ with the constraints $x'\geq x$ and $x'\geq -x$. Minimizing $x'$ ultimately yields the same result as minimizing $|x|$. Going beyond the linear programming and convex optimization paradigm, we can turn to quadratic optimization \cite{nocedal2006quadratic} to incorporate expressions of the form $\sqrt{x}$, with $x\geq 0$, by defining a new variable $x'$ and adding the restriction $x'^2=x$ which can easily be imposed in a quadratic program. With that, we can reformulate an optimization 
of the general expression \eqref{eq:ineqgen} into
\begin{equation}
\begin{aligned}
    &\min_{\bar{x},\bar{y},x,y}\sum_i  \bar{x_i} + \sum_i b_i \bar{y_i}\\
    &\text{ s.t. } \quad -\bar{x}\leq x \leq \bar{x}, \quad \bar{y_i}^2=y_i \quad \forall i,
\end{aligned}
\end{equation}
which is a typical instance of quadratic programming \cite{nocedal2006quadratic}.

\subsection{Observational case}
First, let us recall the definition of the $E_{\pm}$ subspace, as the set of probability distributions,  $S_{\pm}\ni P(a,b,c)$ such that
\begin{equation}
    \label{eq: quantum_corr_pm}
\begin{aligned}
     &\langle A \rangle_1=\langle C \rangle_1=0\\
     &\langle A \rangle _{-1}=-\langle C \rangle_{-1}\\
     &\langle AC \rangle_{-1}=\frac{1}{4}\\
     &\langle B \rangle = -\frac{1}{2}.
\end{aligned}
\end{equation}
where we denominate the remaining free parameters as $\langle AC \rangle_{1}\equiv E_+$ and $\langle C \rangle_{-1}\equiv E_-$. Equivalently, $S_{\pm}$ can be described by the probability distributions 
\begin{equation}
\label{eq:quantump_1}
\begin{aligned}
     &P(a,b=1,c)=\frac{1}{16}\left(1+4ac E_+ \right),\\
     &P(a,b=-1,c)=\frac{1}{16}\left(3+ ac+4\left(c-a\right)E_-\right).
\end{aligned}
\end{equation}
To guarantee trivial constraints, such that $P\geq 0$, we must impose that $|E_+|\leq \frac{1}{4}$ and $|E_-|\leq \frac{1}{4}$.
For the observational case,  we initially have $\mathcal{I}=\sqrt{I_+}+3\sqrt{I_-}$, with $I_+=1+4E_+$ and $I_-=\frac{1+4E_-}{8}$. Then, we consider the QP
\begin{align}
    \max_{E_+,E_-}~ & F_+ + 3F_-\\
    \textrm{s.t.} ~ & P(a,b,c) \in S_{\pm}\cap \Cuc\\
    &F^2_{\pm}=I_{\pm}
\label{eq:obs_problem}
\end{align}
where $\Cuc$ is the set of classical compatible distributions in the UC scenario, shown in Eq. (1) of the main text. It is sufficient to consider that the source $\gamma$ carries the information of how Alice responds to each of Bob's outcomes and similarly for $\alpha$ and Charlie. Therefore, we may take the sources to be of the form $\gamma=\gamma_0\gamma_1$ and $\alpha=\alpha_0\alpha_1$ with a stochastic response function $p(b|\gamma,\alpha)$ for Bob and deterministic response functions $\delta_{a,\gamma_b}$ and $\delta_{c,\alpha_b}$ for Alice and Charlie respectively. This model yields 

\begin{align}
    P(a,b,c) =\sum_{\gamma,\alpha}p(\gamma)p(\alpha)\delta_{a,\gamma_b}\delta_{c,\alpha_b}p(b|\gamma,\alpha)
\label{eq:cuc_constraints}
\end{align}
where $p(\alpha),p(\gamma), p(b|\alpha,\gamma) \geq 0$, $\sum_{\alpha}p(\alpha)=\sum_{\gamma}p(\gamma)=1$ and $\sum_b p(b|\alpha,\gamma)=1$  $\forall \alpha, \gamma$.

To generalize inequality $\mathcal{I}$ for a generic distribution $P$, we must encode the QP constraint $P\in S_{\pm}$ into the objective function. This is done by incorporating penalty functions that guarantee all the constraints of the $E_{\pm}$ subspace. Notably, equations \eqref{eq: quantum_corr_pm} and \eqref{eq:quantump_1} impose
\begin{align}
\label{eq:penalty_constr}
\nonumber
    &P(b=1)=\frac{1}{4}, \quad \sum_{a\neq c}P(a,-1,c)=\frac{1}{4}\\ \nonumber
    &P(i,-1,i)=\frac{1}{4} \quad \forall i, \quad \langle AC\rangle_{-1}=\frac{1}{4}\\ 
    &\langle A\rangle_{-1}=-\langle C\rangle_{-1},\quad \langle A\rangle_1=\langle C\rangle_1=0,
\end{align} 
which can be transformed in the penalty function 
\begin{align*}
    s(P)=&18\left|P(b=1)-\frac{1}{4}\right|+18\left|\sum_{a\neq c}P(a,-1,c)-\frac{1}{4}\right|\\
    &+4\sum_{a=c}\left|P(a,-1,c)-\frac{1}{4}\right| +4\left|\langle AC\rangle_{-1}-\frac{1}{4}\right| \\
    &+|\langle A\rangle_{-1}+\langle C\rangle_{-1}|+|\langle A\rangle_{1}|+|\langle C\rangle_1|,
\end{align*}
where the coefficients here were chosen to try to optimize the robustness of the violation, i.e., they were heuristically changed with the goal to reach an optimal value sufficiently close to the bound considering the $E_{\pm}$ subspace. 

Using quadratic programming with this extra penalty function, one can prove that the following inequality is valid for classical correlations:
\begin{eqnarray}
\label{eq:ineqI_sm}
\nonumber
    \mathcal{I'}=2\sum_{a= c}\sqrt{P(a,1,c)}+3\sqrt{P(-1,-1,1)} -s(P) \\
    \leq \frac{11}{6} + \frac{4\sqrt{3}}{9}.
\end{eqnarray}
Furthermore, using the fact that if we restrict to the subspace of $E_{+}$ and $E_{-}$, it follows that $P(a,1,c)=\frac{1}{16}(1+4E_+)$ for $a=c$ and $P(-1,-1,1)=\frac{1}{8}(1+4E_-)$, one can readily recover the expression $\mathcal{I}$ of the main text.

Using a standard QP optimizer \cite{gurobi}, we consider the problem
\begin{align*}
    \max_{P}~ & \mathcal{I'}\\
    \textrm{s.t.} ~ & P(a,b,c) \in \Cuc\\
\label{eq:obs_problem_2}
\end{align*}
and find that the strategy that maximizes $\mathcal{I'}$ is given by: $\gamma_0,\gamma_1,\alpha_0,\alpha_1$ as bits and \begin{align*}
    p(\gamma) =& 0.35485[00]+0.11556[10]\\
   & +0.16415[01]+0.36544[11]\\
    p(\alpha) = &0.19920[00]+0.33039[10]\\
    &+0.35473[01]+0.11568[11]
\end{align*}
Then for the response function of $B$, $p(b|\gamma,\alpha)$ we have $p(b|\gamma,\alpha)=\delta_{b,0}$ for $(\gamma,\alpha)\in \{( 0 0, 0 1),(1 0, 1 1)\}$, $p(b=0|\gamma = 11,\alpha=10)=0.93183$ and $p(b|\gamma,\alpha)=\delta_{b,1}$ otherwise. This strategy reaches $\mathcal{I'}(P) \approx 2.60311$ and we observe that it satisfies $\mathcal{I'}\leq \frac{11}{6} + \frac{4\sqrt{3}}{9} \approx 2.603133692$ and we obtain this classical upper bound using QP with a precision of $10^{-5}$.

%As mentioned, inequality $\mathcal{I'}$ reduces to inequality $\mathcal{I}$ when restricted to the $E_{\pm}$ subspace by construction, as $s(P)=0$ in this case, and the inequality is given by 
%\begin{equation}
%    2\sum_{a= c}\sqrt{P(a,1,c)}+3\sqrt{P(-1,-1,1)} -s(P)\leq \frac{11}{6} + \frac{4\sqrt{3}}{9}.
%\end{equation}

\subsection{Interventional case}

For the interventional case, we have to incorporate new constraints on the \emph{do-conditional} probabilities generated by performing interventions. This further restricts the $E_{\pm}$ subspace, adding the constraints 
\begin{equation}
\label{eq:do_pm}
    \langle A\rangle_{ \Do{b}}=\langle C\rangle_{\Do{b}}=0,
\end{equation}
where $\langle A \rangle_{ \Do{b}}:=\sum_{a}{a}P(a| \Do{b})$ and $\langle C \rangle_{ \Do{b}}:=\sum_{c}{c}P(c|\Do{b})$. Similarly to the observational case, we have to incorporate these constraints into the objective function via penalty functions. Here, we start by proposing a different penalty function
\begin{align}
    g(P)=&\sum_{a=c}\left|P(a,-1,c)-\frac{1}{4}\right|
        +\left|\langle AC\rangle_{-1}-\frac{1}{4}\right|\\
       & + \left|\sum_{a\neq c}P(a,-1,c)-\frac{1}{4}\right|
        +|\langle A\rangle_{-1}+\langle C\rangle_{-1}|\\
   & +|\langle A\rangle_1|+|\langle C\rangle_1|
        +18\left|P(b=1)-\frac{1}{4}\right|\\
   & +\sum_b\left(|\langle A\rangle_{\Do{b}}|+|\langle C\rangle_{\Do{b}}|\right).
\end{align}
which considers not only the constraints in \eqref{eq:penalty_constr} but also \eqref{eq:do_pm}, and we consider the expression $\mathcal{F}=\sqrt{I_+}+4\sqrt{I_-}$. Using the same arguments as before, we can propose the analogous $\mathcal{F'}$ expression as 
\begin{equation}
    \mathcal{F'}=2\sum_{a= c}\sqrt{P(a,1,c)}+4\sqrt{P(-1,-1,1)} -g(P)
\end{equation}
and consider the QP problem 
\begin{align*}
    \max_{P}~ & \mathcal{F'}\\
     \textrm{s.t.} ~ & \{P(a,b,c), P( a| \Do{b}),P(c | \Do{b})\} \in \Cuc^{\Do{b}}
\label{eq:obs_problem_3}
\end{align*}
where now the requirement of being in $\Cuc^{\Do{b}}$ is imposed on both the interventional and observational distribution, which should be compatible with the UC scenario, that is, besides the constraints in~\eqref{eq:cuc_constraints}, we also have 
\begin{align*}
    P( a| \Do{b})=\sum_{\gamma}p(\gamma)\delta_{a,\gamma_b}\\
    P( c| \Do{b})=\sum_{\alpha}p(\alpha)\delta_{c,\alpha_b}.
\end{align*}

Solving the QP yields $\mathcal{F'}(p) \approx 3.00001$ with optimal solution given by
\begin{align*}
    p(\gamma) =& 0.26845[00]+0.23154[10]\\
    &+0.23154[01]+ 0.26847[11]\\
    p(\alpha) = &0.23163[00]+0.26836[10]\\
    &+0.26836[01]+0.23165[11]
\end{align*}
and $p(b|\gamma,\alpha)=\delta_{b,0}$ for $\gamma,\alpha\in\{(0 0 ,0 1),(1 0 ,1 1)\}$, 
\begin{align*}
    &p(b=0|\gamma=00,\alpha=11)=0.85161,\\
    &p(b=0|\gamma=10,\alpha=01)=0.85223,\\ &p(b=0|\gamma=10,\alpha=10)=0.14812,\\ 
    &p(b=0|\gamma=11,\alpha=11)=0.14801,
\end{align*}
and $p(b|\gamma,\alpha)=\delta_{b,1}$ otherwise. Finally, we observe that 
\begin{equation*}
    \mathcal{F'} \leq \frac{9}{7} + \frac{7}{10}\sqrt{6} \approx 3.000357.
\end{equation*}

\section{Maximum values and robustness of the quantum violation}
\label{app:max_n_rob}
In this section, we discuss in more detail the study of the robustness and maximal values of the quantum violation for $\mathcal{I'}$ and $\mathcal{F'}$. To probe this non-classicality robustness and experimental feasibility, we must consider noise models that try to capture the experimental imperfections of the setup with parameters specified by the success probability of the preparation or measurement procedure. In an entanglement-swapping scenario, different sources of noise must be considered. The main ones
affect state preparation and Bell-state measurement. We consider the same strategy delineated in the main text. We choose $\sigma_x, \sigma_z$ as observables for both A and C with $b=0$ and $b=1$ respectively and, for state preparation, a paradigmatic choice in this context is to introduce a visibility $v$ in each source, given by 
\begin{equation}
    \psi^v_\gamma =\psi^v_{\alpha}=v \ketbra{\psi^{+}(\frac{\pi}{4})}{\psi^{+}(\frac{\pi}{4})}+ (1-v)\frac{\eye}{4},
\end{equation}
where we assume that the sources may prepare the state $ |\psi^{+}(\frac{\pi}{4})\rangle\langle \psi^{+}(\frac{\pi}{4})|$ with probability $v$ and with probability $(1-v)$ it prepares a completely mixed state $\frac{\eye}{4}$.

For an imperfect Bell-state measurement, we consider a noise model inspired by a Bell-state measurement in a Beam Splitter (BS) which strongly relies on the indistinguishability of the photons. Typically, such measurements aim to project into one of the elements of the Bell basis, given by choice of angle $\pi/4$ and, in this case, it has been shown that partial distinguishability in Bell-state measurement acts mixing the $|\psi^{+}\rangle$ results with $|\psi^{-}\rangle$ ones and vice-versa, and similarly for the other elements of the Bell basis $|\phi^{\pm}(\frac{\pi}{4})\rangle$, given by $\ket{\phi^{\pm}(\beta)} = \cos(\beta)\ket{00} \pm \sin(\beta)\ket{11}$,
but it doesn’t mix elements belonging to these two different subspaces. Imperfections in the measurement can be seen as the probability $\frac{1 -\eta}{2}$ of no success, i.e. to read $b=1$ when the real state is in some noise state $\rho_{noise}$. Following the intuition of the BS Bell-state measurement delineated above we can propose $\rho_{noise}(\theta)=\ketbra{\psi^{-}(\theta)}{\psi^{-}(\theta)}$, and the measurement is given by 
\begin{multline}
    E^{\eta}_{b=1}=\left(\frac{1+\eta}{2}\right)\ketbra{\psi^{+}(\theta)}{\psi^{+}(\theta)} +\\+ \left(\frac{1-\eta}{2}\right)\ketbra{\psi^{-}(\theta)}{\psi^{-}(\theta)},
\end{multline}
where $\eta$ was chosen as a parameter of imperfection such that the probability of success is never below 50\% for $0 \leq \eta \leq 1.$ We also have $ E^{\eta}_{b=-1}=\eye - E^{\eta}_{b=1}$, since we only project onto one element of the basis.

These imperfections affect the correlations we may observe in the experiment, yielding probabilities $P_{\theta}^{v,\eta}(a,b,c)$ given by
\begin{equation}
\begin{aligned}
    &P_{\theta}^{v,\eta}(a,1,c)=\frac{1}{16}(1+\eta v^2\sin(2\theta)ac)\\
    &P_{\theta}^{v,\eta}(a,-1,c)=\frac{1}{16}(3+v^2ac+v\cos(2\theta)(c-a))
\end{aligned}
\end{equation}
where $\eta,v\in[0,1]$. This leads to 
\begin{align*}
    \mathcal{I'}(\theta,\eta,v)&=\sqrt{1+\eta v^2\sin(2\theta)} \\
    &+ \frac{3}{4}\sqrt{3-v^2+2v\cos(2\theta)} -15 \frac{|v^2 - 1|}{4}
\end{align*}
and 
\begin{align*}
      \mathcal{F'}(\theta,\eta,v)&=\sqrt{1+\eta v^2\sin(2\theta)} \\
      &+\sqrt{3-v^2+2v\cos(2\theta)} - \frac{|v^2 - 1|}{2}.
\end{align*}

We can explore these expressions numerically to find suitable visibilities $\eta, v$ so that we can still find a violation. For $\eta=1$, these quantities are maximized, for $v=1$, by $\theta_{\mathcal{I'}} = \arctan(2/5)$ and $\theta_{\mathcal{F'}} = \arctan(1/3)$ respectively, which, for perfect visibility $v=1$, gives $\mathcal{I'}_Q(\theta_{\mathcal{I'}},1, 1) \approx 2.69258$ and $\mathcal{F'}_Q(\theta_F, 1,1) \approx 3.16228$, slightly above the violations obtained with $\theta = \pi/8$ presented in the main text. Furthermore, by taking $\theta =\pi/8$ and $v=1$ we can find violation of $\mathcal{I'}$ for $\eta\geq 0.69$, and taking $v=0.99$ we can violation of $\mathcal{I'}$ for $\eta\geq 0.97$. For $v\leq 0.98$ there is no violation of the inequality. In the interventional case, i.e. for the expression $\mathcal{F'}$, we can find better noise robustness for the sources and, for $\theta=\pi/8$, we have for $v=1$ a violation for any $\eta\geq 0.5$, and for $v=0.9$ we find a violation of $\mathcal{F'}$ for $\eta\geq 0.92$. Performing numeric optimization shows that, at least using bipartite 2-qubits states, these values are optimal for these inequalities.

\end{document}